\documentclass[aps,pre,preprint,superscriptaddress,amsmath,amssymb,showpacs]{revtex4}

\usepackage{graphicx}
\usepackage{dcolumn}
\usepackage{bm}

\begin{document}

\title{A functional clustering algorithm for the analysis of dynamic network data}

\author{S. Feldt}
\email{sarahfel@umich.edu}
\affiliation{Department of Physics,
University of Michigan, Ann Arbor, Michigan 48109, USA}

\author{J. Waddell}
\affiliation{Department of Mathematics, University of Michigan, Ann
Arbor, Michigan 48109, USA}

\author{V. L. Hetrick}
\affiliation{Department of Psychology, University of Michigan, Ann
Arbor, Michigan 48109, USA}

\author{J. D. Berke}
\affiliation{Department of Psychology, University of Michigan, Ann
Arbor, Michigan 48109, USA}

\author{M. \.{Z}ochowski}
\affiliation{Department of Physics, University of Michigan, Ann
Arbor, Michigan 48109, USA} \affiliation{Biophysics Program, University of Michigan, Ann Arbor, Michigan 48109, USA}

\date{\today}

\begin{abstract}
We formulate a novel technique for the detection of functional
clusters in discrete event data. The advantage of this algorithm is that no prior knowledge of the number of functional groups is needed, as our procedure progressively combines data traces and
derives the optimal clustering cutoff in a simple and intuitive
manner through the use of surrogate data sets. In order to demonstrate the power of this algorithm to detect changes
in network dynamics and connectivity, we apply it to both simulated neural spike train
data and real neural data obtained from the mouse hippocampus during
exploration and slow-wave sleep. Using the simulated data, we show that our algorithm performs better than existing methods.  In the experimental data, we observe state-dependent
clustering patterns consistent with known neurophysiological
processes involved in memory consolidation.
\end{abstract}


\maketitle

\section{Introduction}
The detection of structural network properties has been recently recognized to be of great importance in aiding understanding of the properties of variety of man-made and natural networks \cite{strogatz2001,albert2002,newman2003,schwarz2008}.  Here, however, two significantly different notions of network structure have to be identified. One is the physical (or anatomical) structure of the network.  In this case, community structure refers to groups of nodes within a network which are more highly connected to other nodes in the group than to the rest of the network. Here, multiple techniques exist which utilize a knowledge of the network topology (adjacency matrix) to extract this hidden structure \cite{girvan2002,leicht2008,newman2006,danon2005}.

The other type of structure is the functional structure, which refers to a commonality of function of subsets of units within the network, generally observed by monitoring the similarities in the dynamics of nodes \cite{fingelkurts2005,friston1993}. Thus the structural proximity (i.e existence of physical connection between the network elements) is replaced with notion of functional commonality (or proximity), which can rapidly evolve based on the observed dynamics.

While the physical connectivity of the network can be obtained for many man-made and some biological networks, there are large classes of networks where physical or anatomical structure can not be obtained. The brain is a prime example of such a system - the cortex alone contains around $1.5\times 10^{14}$ tightly packed connections (synapses) and it is clearly impossible to derive any detailed properties of its connectivity. It is not even completely clear that having such a detailed knowledge of the connectivity would be particularly useful in understanding brain function, as it significantly evolves during the life-time of an individual through such processes as neuronal loss, adult neurogenesis and constant rewiring (i.e. creation, annihilation and modulation of synapses). Also, since it is known that brain function is distributed over large neuronal ensembles, or even more globally, between different brain modalities, it becomes imperative to understand how these ensembles self-organize to generate desired functions (movement, memory storage/recall, etc.) \cite{hebb1949,gerstein1978,singer1999,zhou2007}.
The advent of techniques that allow the activity of many cells to be simultaneously monitored provides hope for a clearer understanding
of these neural codes, but also demands novel tools for the
detection and characterization of spatio-temporal patterning of this
activity.

While it is assumed that these ensembles are formed dynamically \cite{milner1974,engel2001,vondermalsburg1995,singer2001}, through spatio-temporal interactions of activity patterns of many individual neurons, the neural correlates of cognition are not well understood. One of the most prominent hypothesis addressing this issue is the temporal correlation hypothesis \cite{vondermalsburg1981,gray1999,singer1993,engel1991}. Namely, it is assumed that correlations between activity pattern of neurons mediate feature binding and thus formation of intermittent functional ensembles in the brain. Thus, functional clustering can potentially be reduced to the identification of temporally correlated groups of neurons.

In order to successfully capture the (physical or functional) community structure of a network, a clustering algorithm should have two important properties:  the ability to detect relationships between nodes in order to form clusters, and the ability to determine the specific set of clusters which optimally characterize the network structure. While some clustering methods have been designed to extract the structure directly from the dynamics of the neurons \cite{gerstein1985,baker2000,dayhoff1994,lindsey2006,gerstein1978}, most methods rely on using a similarity measure to compute distances in similarity space between neurons, and then use structural clustering methods to determine the functional groupings \cite{borgatti1994,boccaletti2006,berger2007,kreuz2007,ozden2008}. However, a major problem becomes identifying statistically significant community structures from spurious ones. To achieve this goal, current structural clustering techniques  involve an optimization of the network modularity \cite{newman2004a,newman2004b} or require a prior knowledge of the number of communities \cite{newman2003,girvan2002,fortunato2004,zhou2003,newman2004c,ball1967}.

In this paper, we develop a novel clustering method that does not depend on structural network information, but instead derives the functional network structure from the temporal interdependencies of its elements. We refer to this method as the
Functional Clustering Algorithm (FCA). The key advantage of this algorithm is that it incorporates a natural cutoff point to cease clustering and obtain the functional groupings without an a priori knowledge of the number of groups.  Additionally, the algorithm can be used with a variety of different similarity measures, allowing it to detect functional groupings based on multiple features of the data.
Although this paper focuses on the application to neural data in the form of spike trains, the FCA can be applied to any type of discrete event data.

The paper is organized as follows: we first introduce the Functional Clustering Algorithm, along with a novel similarity metric designed to detect co-firing events in neural data.  We then compare the performance of the algorithm to two existing methods using simulated data, and show that it performs better than existing measures. Finally, we demonstrate the application of our new algorithm to experimental data exploring progressive memory consolidation in the hippocampus.

\section{The Functional Clustering Algorithm}
Here we introduce the Functional Clustering Algorithm (FCA) which is tailored to detect functional clusters of network elements.  The algorithm can be applied to any type of discrete event data, however, this paper will focus only on the application of the algorithm to neural spike train data.

The FCA dynamically groups pairs of spike trains based on a chosen similarity metric, forming progressively more complex spike patterns.  We will introduce a new similarity measure which is used for the data analyzed in this paper, but any pairwise similarity measure can be chosen. The specific choice of the metric should depend on the nature of the data being analyzed and the type of functional relationships which one chooses to detect.

\begin{figure}
\includegraphics{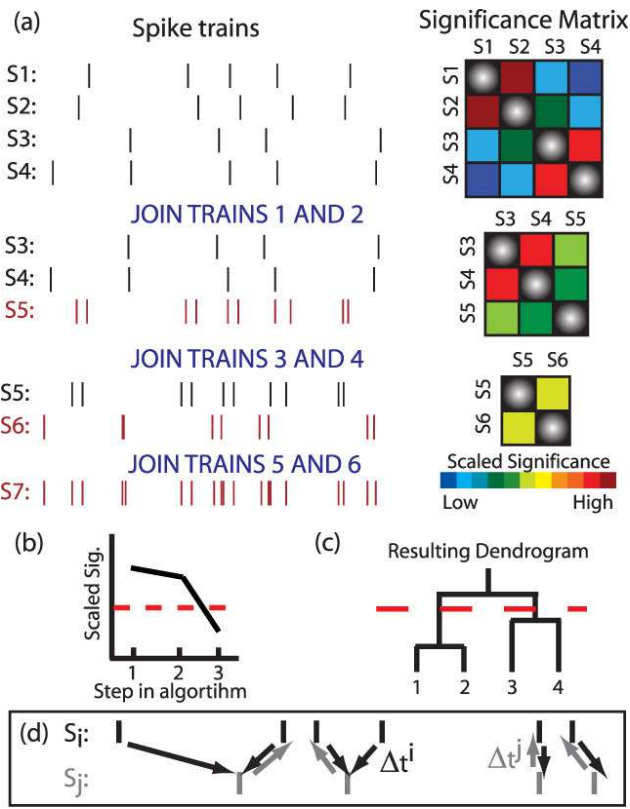}
\caption{\label{fig1} (Color online) Functional Clustering
Algorithm.  (a) An example of the algorithm applied to four spike
trains.  Two trains are merged in each step by selecting the pair of
neurons with the highest scaled significance value and effectively creating a new neuron
by temporally summing their spike trains.  The procedure is repeated
until one (complex) spike train remains.  (b) We cease clustering when the trains being grouped are no longer significant; here the dotted red line denotes the significance cutoff.  (c) The subsequent dendrogram
obtained from the FCA.  The dotted line
denotes the clustering cutoff.  (d) Schematic of the average minimum distance between
spike trains.}
\end{figure}

A general description of the FCA is as follows (see the subsequent sections for detailed descriptions and Fig. \ref{fig1} for a schematic of the algorithm):
\begin{list}{}{}
\item[1.] We first create a matrix of pairwise similarity values between all spike trains.
\item[2.] We then use surrogate data sets to calculate $95\%$ confidence intervals for each pairwise similarity.  These significance levels are used to calculate the scaled significance between each pair of similarity values (see Sect. \ref{sig} for the definition of scaled significance).
\item[3.] The pair of trains with the highest significance is then chosen to be grouped together, and the scaled significance of this pair is recorded. A unique element of the FCA is that the two spike trains which are grouped together are then merged by joining the spikes into a single new train (see Fig. \ref{fig1}(a)).  This allows for a cumulative assessment of similarity between the existing complex cluster and the other trains.
\item[4.]  The trains which are being joined are then removed, the similarity matrix is recalculated for the new set of trains, new surrogate data sets are created, and a new scaled significance matrix is calculated.
\item[5.] We repeat the joining steps ($3-4$), recording the scaled significance value used in each step of the algorithm until the point at which no pairwise similarity is statistically significant, indicating that the next joining step is not statistically meaningful.  We refer to this step as the clustering cutoff (dashed red line in Fig. \ref{fig1}).  At this point, the functional groupings are determined by observing which spike trains have been combined during the clustering algorithm.
\end{list}

A key advantage of this algorithm is that the ongoing comparison of the similarity metric obtained from the data with that from the surrogates causes the algorithm to have a natural stopping point, meaning that one does not need an \emph{a priori} knowledge of the number of functional groups embedded in the data.  Gerstein et al \cite{gerstein1978} also developed an aggregation method based on
grouping neurons with significant coincident firings, but this method results in the formation of non-unique strings of related neurons as opposed to well defined functional groupings.  We now discuss the details of the implementation of the FCA in the following sections.

\subsection{Average Minimum Distance}
For the data presented in this paper, we use a new similarity metric which we call the Average Minimum Distance (AMD) to determine functional groupings.  The AMD is useful in capturing similarities due to coincident firing between neurons.  Note that other metrics could be chosen, depending upon the nature of the recorded data.  To compute the AMD between two spike trains $S_i$ and $S_j$, we calculate the distance $\Delta t_k^i$ from each spike in $S_i$ to the closest spike in $S_j$ as shown in Fig. \ref{fig1}(d).  We then define
\begin{equation}
D_{ij/ji}=\frac{1}{N_{i/j}}\sum_k \Delta t^{i/j}_k
\end{equation}
where $N_{i/j}$ is the total number of spikes in $S_i$ or $S_j$ respectively.  Finally, we define the AMD between spike trains $S_i$ and $S_j$ to be
\begin{equation}
AMD_{ij}=\frac{D_{ij}+D_{ji}}{2}.
\end{equation}

\subsection{Calculation of Significance}
\label{sig}

In order to determine the significance between two trains, we create 5,000-10,000 surrogate data sets and calculate pairwise similarities for each surrogate set.  The surrogate spike trains are created by adding a jitter to each spike in the train.  This jitter is drawn from a normal distribution \cite{rolston2007}, similar to the technique developed by Date \cite{date1998}.  The method of adding jitter to spikes (also known as dithering or teetering) to create surrogate data sets is commonly used when analyzing neural data and has been shown to eliminate correlations between spike timings \cite{shmiel2006,pazienti2007}. Creating the surrogate trains in this manner preserves the frequency of each train while keeping the gross properties of the interspike-interval distribution.

We examine the distribution of similarity values and create the cumulative distribution function (CDF) to determine the $95\%$ level of significance.  The scaled significance (Fig. \ref{fig2} and Fig. \ref{fig7}) is measured in units defined as the distance from the midpoint of the CDF to the $95\%$ significance cutoff.  Thus, a scaled significance value equal to one denotes the $95\%$ significance level, and values higher than one are significant while values lower than one are deemed insignificant.

\section{Comparison to Other Algorithms}

\begin{figure*}
\includegraphics{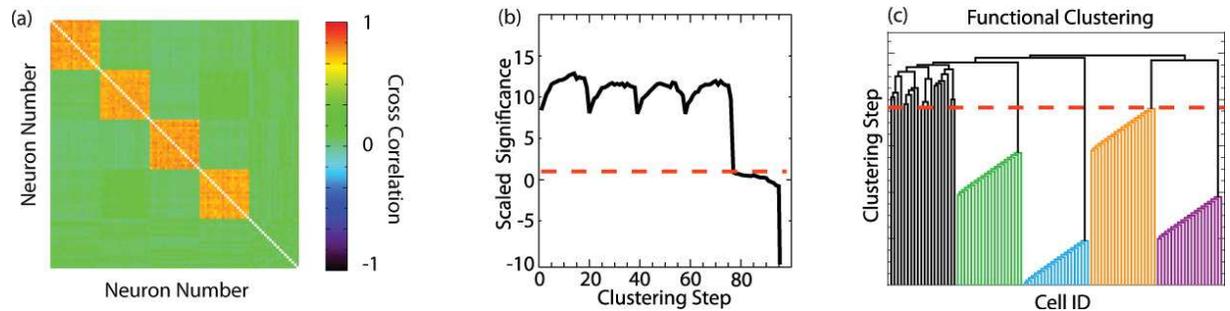}
\caption{\label{fig2}(Color online) Performance of the FCA on simulated data. (a) The cross-correlation matrix showing the correlation structure of the simulated data.  (b) The scaled significance used in each step of the
FCA.  The dashed red line denotes the
point at below which clustering is no longer significant.  ((c) Dendrogram resulting from
Functional Clustering.  The algorithm easily
identifies the correct groups.}
\end{figure*}

In order to verify the performance of the FCA and compare it to that of existing clustering methods, we created simulated spike trains with a known correlation structure.  Specifically, we created a set of 100 spike trains derived from a Poisson distribution that consist of four independent groups, $20$ spike trains each, and 20 uncorrelated spike trains. The spike trains within these four groups are correlated (see Fig. \ref{fig2}(a)).  To create the correlated groups, we first created a master spike train and used this train to create new trains by randomly deleting spikes from the master train with a certain probability.  Thus, the resulting train was also a Poisson process with a firing rate dependent upon the deletion probability.  The master train was 5000 time steps long, with each neuron spiking an average of 250 times during the duration of the train.  To further randomize the timings of the spikes copied from the master train, we added  jitter (drawn from a standard normal distribution) to the spike times.  Each correlated group was composed of 20 trains from the same master. The firing rate of the independent trains was set to match that of the correlated trains.

We first applied the FCA to the simulated data described above (Fig. \ref{fig2}(b-c)).  In Fig. \ref{fig2}(b) we show the scaled significance at each joining step in the algorithm.  The dashed red line marks the significance cutoff (single 95\% confidence interval); points above this line are statistically significant, and the clustering cutoff is given by the point where the curve drops below this line.  Fig. \ref{fig2}(c) shows the resulting dendrogram with the dashed red line denoting the clustering cutoff. The algorithm correctly identifies the 4 groups of neurons as well as the 20 independent neurons.

\subsection{Comparison to the Gravitational Method}
We then compared the performance of the FCA to that of the gravitational method \cite{gerstein1985,baker2000,dayhoff1994,lindsey2006}.  This method performs clustering based on the spike times of neuronal firings by mapping the neurons as particles in N-dimensional space, and allowing their positions to aggregate in time as a function of their firing patterns.  Particles are initially located along the trace of the N-dimensional space and given a `charge' which is a function of the firing pattern on the neuron.  The charge $q_i$ on a particle is given by
\begin{equation}
q_i\left(t\right)=\sum_kK\left(t-T_k\right)-\lambda_i
\end{equation}
where $K(t)=exp\left(-t/\tau\right)$ for $t>0$ and $K=0$ otherwise, $T_k$ are the firing times of the neuron, and $\lambda_i$ is the firing rate of the neuron, normalized so that the mean charge on a particle is zero. The position vector, $\textbf{x}$, of the particle is then allowed to evolve based upon the following rule:
\begin{equation}
\textbf{x}_i\left(t+dt\right)=\textbf{x}_i\left(t\right)+\kappa dt\sum_{j\ne i}q_iq_j\frac{\textbf{x}_j-\textbf{x}_i}{|\textbf{x}_j-\textbf{x}_i|}
\end{equation}
where $\kappa$ is a user defined parameter that controls the speed of aggregation.  One then calculates the Euclidean distance between particles as a function of time and looks for particles which cluster in the N-dimensional space (i.e., the distance between the particles becomes small).

\begin{figure}
\includegraphics{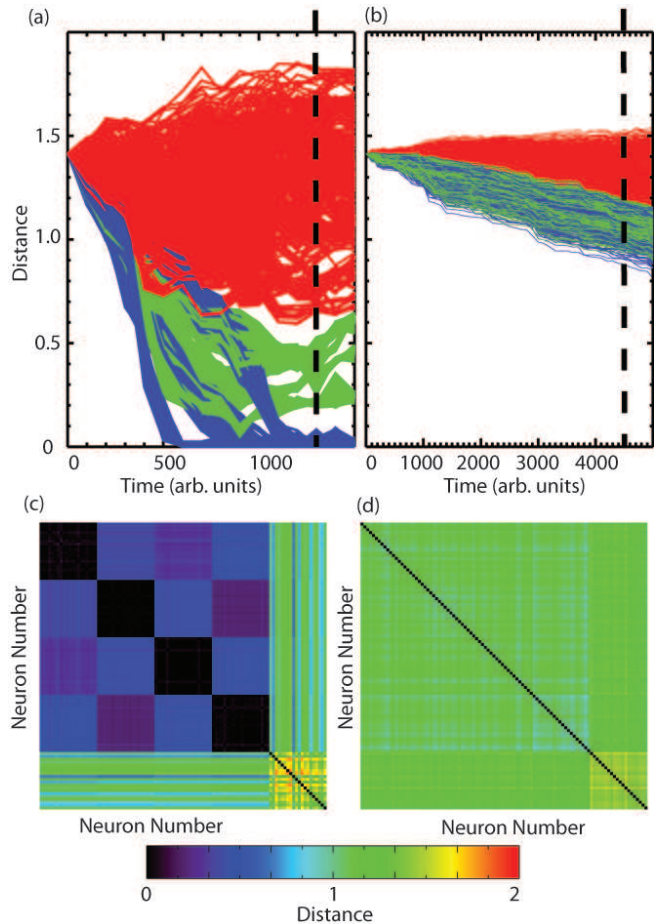}
\caption{\label{fig3}(Color online) Application of the gravitational method to simulated data.  (a,b) Pairwise distances as a function of time in the stimulation for high correlation within clusters (a) and low correlation within clusters (b).  Blue traces: intra-cluster distances, green traces: inter-cluster distances, red traces: distances between any train and an independent train. (c,d) Matrix version of distances for the point in time denoted by the dashed vertical line in (a),(b) respectively.  Note that for the low correlation case, one can not detect the formation of individual clusters.}
\end{figure}

Fig. \ref{fig3} depicts the results of applying the gravitational method to the simulated data described above for cases of high correlation ($C\approx 0.63$) within groups (Fig. \ref{fig3}(a,c)) and also for low correlation ($C\approx 0.13$) within clusters (Fig. \ref{fig3}(b,d)).  In Fig. \ref{fig3}(a-b) we plot the pairwise distances between particles as a function of time in the algorithm.  Blue traces denote distances between intra-cluster trains, green between inter-cluster ones, and red between any train and an independent train.  To visualize the results of the method, we have sliced these plots as indicated by the dashed vertical line and represent the distances at this point in time as matrices in Fig. \ref{fig3}(c-d).  While, for the case of high correlation between the spike trains, the algorithm  separates the 4 groups correctly (black squares in Fig. \ref{fig3}(c)), one is unable to distinguish between inter and intra-cluster trains for the low correlation case.  Furthermore, these plots must be visually inspected for the cutoff (i.e. time point at which they stabilize) and the clustering results may significantly  depend on its position, as the algorithm has no inherent stopping point and the rate of aggregation is parameter dependent.  Even then, the detection of the formed clusters may require the application of an additional N-dimensional clustering algorithm to detect the clusters formed in the N-dimensional space.  Another drawback of this method is that as the particles aggregate into clusters, the clusters start interacting due to the nature of the algorithm, causing inter-cluster distances to become significantly lower than those with random trains, which does not match the correlation structure of the data.

The FCA performed the correct clustering of the data for the case of the high correlation and only made an occasional error for data with the low correlation.

\subsection{Comparison to Complete Linkage and Modularity}
We next compare the performance of the FCA to a method which maps spiking dynamics onto a structural space and then uses a structural clustering method to determine functional groupings.  The structural clustering method used is a standard hierarchical clustering technique called complete linkage.  Since this algorithm has no inherent cutoff point at which clustering is stopped, we combine it with a calculation of the weighted modularity \cite{newman2004b}, which is a commonly used measure to determine the best set of groupings when dealing with hierarchical clustering methods. We have also tried other methods (single-linkage, GN algorithm  \cite{borgatti1994,girvan2002}), but complete linkage gave the best results of the other methods attempted.  Please see \cite{boccaletti2006,borgatti1994} for a review of standard hierarchical clustering techniques.

The complete linkage algorithm again clusters trains based upon a similarity measure.  In this algorithm, a similarity matrix is created and the elements with the maximum similarity are joined.  However, the clusters are formed through virtual grouping of the elements and there is no re-calculation of the similarity measure; the similarity between clusters is simply defined to be the minimum similarity between elements of the clusters. For the data presented in this paper, we use the absolute value of the normalized cross-correlation matrix as our similarity matrix, since this is what is commonly used to do examine community structure in neuroscience applications.  To compute this matrix, spike trains are first convolved with a gaussian kernel and the signal is demeaned (the mean value of the signal is subtracted).  The cross-correlation is given by
\begin{equation}
\hat{C}\left(S_i,S_j\right)=\left|\frac{C\left(S_i,S_j\right)}{\sqrt{C\left(S_i,S_i\right)\cdot
C\left(S_j,S_j\right)}}\right|,
\end{equation}
where $C$ is the linear cross correlation function
\begin{equation}
C\left(S_i,S_j\right)=\int_{-\infty}^{\infty}S_i\left(t\right)S_j\left(t\right)dt.
\end{equation}

Since the complete linkage algorithm has no inherent method of determining the clustering cutoff, we compute the (weighted) modularity \cite{newman2004b} for each step of the algorithm.  The modularity measure was originally tailored to detect the optimal community structure based upon structural connections between nodes (i.e. adjacency matrix), however it can also be used to detect optimal clustering based on not structural, but dynamical relations, where the adjacency matrix is substituted with the correlation matrix. The modularity is given by
\begin{equation}
Q=\frac{1}{2m}\sum_{ij}\left(A_{ij}-\frac{k_i k_j}{2m}\right)\delta\left(c_i,c_j\right)
\end{equation}
where $A_{ij}$ is our similarity matrix, $k_i=\sum_j A_{ij}$, $m=\frac{1}{2}\sum_{ij}A_{ij}$, and $\delta \left(c_i,c_j\right)=1$ if $i$ and $j$ are in the same community and zero otherwise.  The maximum value of the modularity is then used to define the clustering cutoff.

\begin{figure}
\includegraphics{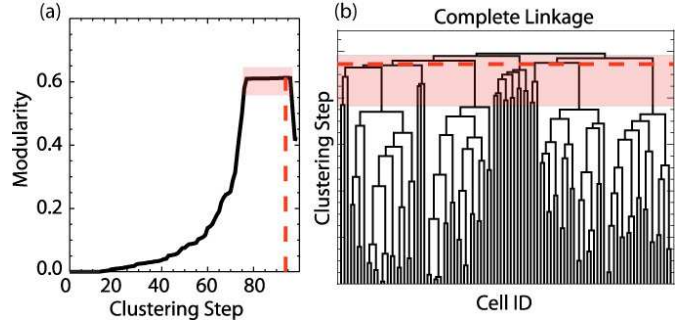}
\caption{\label{fig4}(Color online) (a) Modularity
calculation for the clustering obtained using complete linkage. The
transparent red box marks the ambiguous cutoff area. (b) Dendrogram indicating
clustering by complete linkage. Here the clustering cutoff is ambiguous and the algorithm fails to identify the appropriate structure.}
\end{figure}

The complete linkage dendrogram is shown in Fig. \ref{fig4}(b) and the modularity for this clustering is plotted in Fig. \ref{fig4}(a). The clustering cutoff is defined as the maximum of the
modularity \cite{newman2004a,newman2004b}, however the scaling of the
modularity, even in this simple case, provides ambiguous results. The
numerical maximum of the modularity is observed for the clustering
step marked by the dashed red line in Fig. \ref{fig4} - significantly
above the clustering step that starts linking random spike trains.
Even if we relax this definition and assume that the set of high
modularity values is equivalent, the exact location of the cutoff
is ambiguous as shown by the area enclosed in the transparent red
box. Note that the FCA does not have this ambiguity, as the
cutoff is quite clear and the algorithm correctly identifies the
groups embedded in the spike train data.

\begin{figure}
\includegraphics{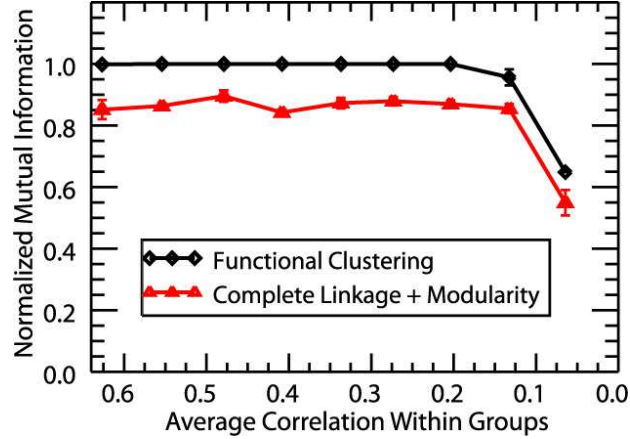}
\caption{\label{fig5}(Color online) Normalized mutual information as a function of average group correlation.  The measure takes a maximal value of one when the established clustering structure matches the predetermined groups and $NMI\rightarrow 0$ when the obtained clustering structure is independent of the original groupings.  Functional Clustering identifies the correct group structure for almost all values of correlation while complete linkage and modularity consistently create erroneous structure.}
\end{figure}

To further explore the performance of the FCA in comparison with complete linkage and modularity, we monitor the performance of both methods for progressively lower correlations within the four clustered groups (Fig. \ref{fig5}). We did not perform this analysis for the gravitational method since that algorithm has no predetermined stopping point and cluster identification must be assessed by the user.  As before, the inter-cluster correlation is controlled through progressive, random deletion of spikes from a master train. In order to compare the performance of the two algorithms, it is necessary to compare the obtained clusterings to the known structure of the data. To assess the correctness of the retrieved clusters as compared to the actual structure of the network, we calculate the normalized mutual information (NMI) \cite{fred2003,danon2005} as a function of the average correlation within the constructed groups.  The NMI is a measure used to evaluate clustering algorithms and determine how well the obtained clustering, $C'$, matches the original structure, $C$.  To compute the NMI, one first creates a matrix with $c$ rows and $c'$ columns, where $c$ is the number of communities in $C$ and $c'$ is the number of found communities in $C'$.  An entry, $N_{ij}$, is defined to be the number of nodes in community $i$ that have been assigned to the found community $j$.  If we denote $N_{i/j}=\sum_{j/i} N_{ij}$ and $N=\sum_{ij} N_{ij}$ then we can define
\begin{equation}
NMI\left(C,C'\right)=\frac{-2\sum_{i}\sum_{j}N_{ij}ln\left(\frac{N_{ij}N}{N_iN_j}\right)}{\sum_{i}N_iln\left(\frac{N_i}{N}\right)+\sum_{j}N_jln\left(\frac{N_j}{N}\right)}.
\end{equation}
This measure is based on how much information is gained about $C$ given the knowledge of $C'$.  It takes a minimum value of 0 when $C$ and $C'$ are independent, and a maximal value of 1 when they are identical.

In Fig. \ref{fig5} we use the NMI to compare the obtained clustering with the known structure of the simulated data. As shown in the figure, complete linkage and modularity consistently fail to identify the correct structure.  This is because the maximum of the modularity occurs for a point in the algorithm where various independent spike trains have been joined, creating erroneous group structure.  However, the FCA correctly identifies neurons for almost all values of correlation. Please note that the $80\%$ level of correctness using complete linkage and modularity for higher intercluster correlation values is due to the fact that we had only 24 independent groups (20 spike trains + 4 independent clusters) in the tested network. A higher number of independent neurons would lead to a poorer performance of that method (due to the erroneous grouping of independent neurons) and thus higher relative effectiveness of the FCA.


\section{Application to Experimental Data}

\begin{figure}
\includegraphics{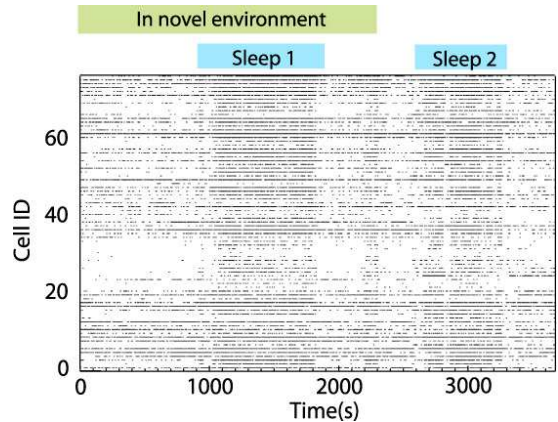}
\caption{\label{fig6}(Color online)  Raster plot of neural data
obtained from an unrestrained mouse during exploration of a novel
environment and sleep.}
\end{figure}

In order to show possible applications of the FCA to
real data, we examined spike trains recorded from the hippocampus of
a freely moving mouse, using tetrode recording methods
\cite{berke2008}.  All animal experiments were approved by the University of Michigan Committee on the Use
and Care of Animals.  In this report, we focus on the population
of pyramidal neurons (77 total; by subregion: 42 CA1,
21 CA2, 14 CA3).  While recording this cell population, the mouse was placed in a
novel rectangular track environment. The mouse initially explored the
environment by running approximately 20 laps, then settled down, and
shortly thereafter fell asleep. A raster plot of this data is shown in Fig. \ref{fig6}.  This data set is of interest for two
reasons. Firstly, there are established differences in the
functional organization of hippocampal networks between active
exploration and slow-wave sleep \cite{buzsaki2003}. These include
the joint activation of pyramidal cell ensembles at
timescales corresponding to gamma frequencies during awake
movement \cite{harris2003}, and the high speed replay of pyramidal
cell sequences within ripple events that occur preferentially during slow-wave sleep and rest \cite{foster2006}. Secondly, the mouse learned a new spatial
representation during exploration of the novel environment (as
indicated by the formation of ``place fields'' \cite{berke2008}) and
the subsequent epoch of slow-wave sleep has been hypothesized to be a period
of memory consolidation \cite{buzsaki1998,kudrimoti1999},
that is presumed to involve alterations in structural and thus functional network connectivity. These structural alterations involve the strengthening of existing monosynaptic connections between the neurons. Furthermore, recent experimental findings have shown that memory consolidation of the neural representation of novel stimuli results in two changes: neurons that are correlated during initial exposure progressively increase their co-firing, while the neurons that have shown a loose relation become further de-correlated  \cite{oneill2008}. In terms of network reorganization, this should lead to the tightening of the cluster of cells involved in the coding of the new environment and, at the same time, a functional decoupling from the other cells.

Given these functional differences between the various behavioral states of the mouse, we expected to see different clustering
patterns during the exploration and sleep phases, due to the known differences in network dynamics between these behavioral states.

\begin{figure}
\includegraphics{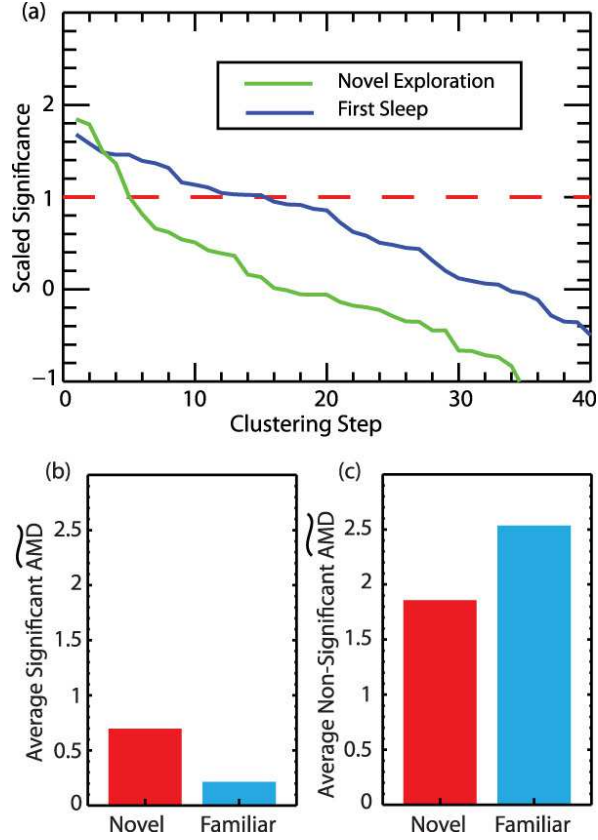}
\caption{\label{fig7}(Color online) (a) The scaled significance used in clustering calculated for novel exploration $(0-200s)$ and the first sleep period $(900-1100s)$.  The significance cutoff is shown by the dashed line.  The FCA is able to detect the greater number of neurons involved in joint firing known to occur during sleep. (b) Comparison of the $\widetilde{AMD}$ averaged over significant clustering steps from novel exploration and a subsequent familiar exploration.  We observe a decrease in this value during the familiar exploration as correlations between neurons become tighter. (c)  Comparison of the $\widetilde{AMD}$ distances averaged over non-significant clustering steps during novel and familiar exploration.  Here we see an increase in this value during familiar exploration as neurons which were uncorrelated become further de-correlated.  }
\end{figure}

In Fig. \ref{fig7}(a) we show the scaled significance used in the FCA during the initial exploration as well as the first sleep period.  For this data, the jitter amount added to surrogates was drawn from a normal distribution with a standard deviation of 10s to destroy long term correlations between neurons. The cutoff point in the algorithm occurs when the scaled significance drops below the dashed red line.  The step in the algorithm at which this cutoff occurs indicates the number of neurons involved in the clustering.  Thus if a cutoff occurs for a late (as opposed to early) step in the clustering, more neurons are recruited into the clusters. One can see that there is an increase in the number of significant pairs being clustered during the sleep period (due to the later stage of cutoff), consistent with the increased co-activation of neurons known to occur during sleep ripples.

We then compared the initial exploration of the novel environment to a subsequent exploration of the same environment (after the sleep epochs).   Here, we hypothesized that, due to memory consolidation and the associated changes in correlations between neurons, we would observe a selective drop in the joining AMD when comparing the initial exposure to a novel environment to a subsequent exposure once the environment has become familiar. This drop should occur for initially small AMD values (initially correlated neurons) as these neurons become further correlated. However, for initially large (insignificant) AMD values, we expect an increase in the AMD values when comparing novel and familiar exploration. This growth occurs as the neurons with low correlations become further uncorrelated.

To assess any changes in the AMD values between initial (novel) and familiar exploration, we must introduce a frequency correction during the calculation of the $D_{ij}$ values (note that the effect of spiking frequency in the measure is accounted for in the algorithm through the comparison to surrogate data).  Here, we normalize these distances by the average expected distance obtained from uniformly distributed spike trains having the same spike frequency:  $D_{ij/ji}^{unif}=(\Delta T)/(N_{j/i}+1)$, where $\Delta T$ is the train length.  Thus,
\begin{equation}
\tilde{D}_{ij/ji}=\frac{D_{ij/ji}}{D_{ij/ji}^{unif}}.
\end{equation}
We then define the $\widetilde{AMD}$ between trains $S_i$ and $S_j$ to be
\begin{equation}
\label{freq_amd}
\widetilde{AMD}_{ij}=\frac{\tilde{D}_{ij}+\tilde{D}_{ji}}{2}.
\end{equation}
Lower values of $\widetilde{AMD}$ indicate tighter functional clustering between the cells.

 In Fig. \ref{fig7}(b), we show changes of the average $\widetilde{AMD}$s used to cluster the
neurons for the clustering steps which have a significantly lower $\widetilde{AMD}$ than that obtained from surrogates (i.e. co-firing cells), during novel exploration and a subsequent familiar exploration. We indeed see that the average $\widetilde{AMD}$ value is lower for neurons during the familiar exploration indicating that the firing patterns of the neurons are more tightly correlated.  Thus, as in the case of \cite{oneill2008}, the observed decrease of the $\widetilde{AMD}$ during the subsequent presentation of the novel environment occurs for neurons which fire in the same spatial locations of the maze. In Fig. \ref{fig7}(c), we show the average $\widetilde{AMD}$ distances for the non-significant clustering steps during the novel and familiar exploration. These distances are greater during the familiar exploration as the activity of the neurons having low correlation becomes even less correlated.

\section{Conclusions}
In conclusion, we have developed a new Functional Clustering
Algorithm to perform grouping based on relative activity
patterns of discrete event data sets. We applied this algorithm to neural spike train data, and have shown that the new algorithm performs better than
existing ones in simple test cases, using simulated data.  Additionally, we showed that the algorithm successfully detects
state-related changes in the functional connectivity of the mouse
hippocampus. Functional Clustering should therefore be a useful tool
for the detection and analysis of neuronal network changes occurring
during cognitive processes and brain disorders, as well other dynamical biological/physical phenomena that can be represented by discrete time series.

\begin{acknowledgments}
This work was funded through an NSF Graduate Research Fellowship (SF), NIH Grant EB008163 (MZ), the Whitehall Foundation (JB), and the National Institute on Drug Abuse R01 DA14318 (JB).
\end{acknowledgments}


\begin{thebibliography}{49}
\expandafter\ifx\csname natexlab\endcsname\relax\def\natexlab#1{#1}\fi
\expandafter\ifx\csname bibnamefont\endcsname\relax
  \def\bibnamefont#1{#1}\fi
\expandafter\ifx\csname bibfnamefont\endcsname\relax
  \def\bibfnamefont#1{#1}\fi
\expandafter\ifx\csname citenamefont\endcsname\relax
  \def\citenamefont#1{#1}\fi
\expandafter\ifx\csname url\endcsname\relax
  \def\url#1{\texttt{#1}}\fi
\expandafter\ifx\csname urlprefix\endcsname\relax\def\urlprefix{URL }\fi
\providecommand{\bibinfo}[2]{#2}
\providecommand{\eprint}[2][]{\url{#2}}

\bibitem[{\citenamefont{Strogatz}(2001)}]{strogatz2001}
\bibinfo{author}{\bibfnamefont{S.~H.} \bibnamefont{Strogatz}},
  \bibinfo{journal}{Nature} \textbf{\bibinfo{volume}{410}},
  \bibinfo{pages}{268} (\bibinfo{year}{2001}).

\bibitem[{\citenamefont{Albert and Barabasi}(2002)}]{albert2002}
\bibinfo{author}{\bibfnamefont{R.}~\bibnamefont{Albert}} \bibnamefont{and}
  \bibinfo{author}{\bibfnamefont{A.~L.} \bibnamefont{Barabasi}},
  \bibinfo{journal}{Reviews of Modern Physics} \textbf{\bibinfo{volume}{74}},
  \bibinfo{pages}{47} (\bibinfo{year}{2002}).

\bibitem[{\citenamefont{Newman}(2003)}]{newman2003}
\bibinfo{author}{\bibfnamefont{M.~E.~J.} \bibnamefont{Newman}},
  \bibinfo{journal}{Siam Review} \textbf{\bibinfo{volume}{45}},
  \bibinfo{pages}{167} (\bibinfo{year}{2003}).

\bibitem[{\citenamefont{Schwarz et~al.}(2008)\citenamefont{Schwarz, Gozzi, and
  Bifone}}]{schwarz2008}
\bibinfo{author}{\bibfnamefont{A.~J.} \bibnamefont{Schwarz}},
  \bibinfo{author}{\bibfnamefont{A.}~\bibnamefont{Gozzi}}, \bibnamefont{and}
  \bibinfo{author}{\bibfnamefont{A.}~\bibnamefont{Bifone}},
  \bibinfo{journal}{Magnetic Resonance Imaging} \textbf{\bibinfo{volume}{26}},
  \bibinfo{pages}{914} (\bibinfo{year}{2008}).

\bibitem[{\citenamefont{Girvan and Newman}(2002)}]{girvan2002}
\bibinfo{author}{\bibfnamefont{M.}~\bibnamefont{Girvan}} \bibnamefont{and}
  \bibinfo{author}{\bibfnamefont{M.}~\bibnamefont{Newman}},
  \bibinfo{journal}{PNAS} \textbf{\bibinfo{volume}{99}}, \bibinfo{pages}{7821}
  (\bibinfo{year}{2002}).

\bibitem[{\citenamefont{Leicht and Newman}(2008)}]{leicht2008}
\bibinfo{author}{\bibfnamefont{E.~A.} \bibnamefont{Leicht}} \bibnamefont{and}
  \bibinfo{author}{\bibfnamefont{M.~E.~J.} \bibnamefont{Newman}},
  \bibinfo{journal}{Physical Review Letters} \textbf{\bibinfo{volume}{100}},
  (\bibinfo{year}{2008}).

\bibitem[{\citenamefont{Newman}(2006)}]{newman2006}
\bibinfo{author}{\bibfnamefont{M.~E.~J.} \bibnamefont{Newman}},
  \bibinfo{journal}{Proceedings of the National Academy of Sciences of the
  United States of America} \textbf{\bibinfo{volume}{103}},
  \bibinfo{pages}{8577} (\bibinfo{year}{2006}).

\bibitem[{\citenamefont{Danon et~al.}(2005)\citenamefont{Danon, Diaz-Guilera,
  Duch, and Arenas}}]{danon2005}
\bibinfo{author}{\bibfnamefont{L.}~\bibnamefont{Danon}},
  \bibinfo{author}{\bibfnamefont{A.}~\bibnamefont{Diaz-Guilera}},
  \bibinfo{author}{\bibfnamefont{J.}~\bibnamefont{Duch}}, \bibnamefont{and}
  \bibinfo{author}{\bibfnamefont{A.}~\bibnamefont{Arenas}},
  \bibinfo{journal}{Journal of Statistical Mechanics} pp.~\bibinfo{pages}{--}
  (\bibinfo{year}{2005}).

\bibitem[{\citenamefont{Fingelkurts et~al.}(2005)\citenamefont{Fingelkurts,
  Fingelkurts, and Kahkonen}}]{fingelkurts2005}
\bibinfo{author}{\bibfnamefont{A.~A.} \bibnamefont{Fingelkurts}},
  \bibinfo{author}{\bibfnamefont{A.~A.} \bibnamefont{Fingelkurts}},
  \bibnamefont{and} \bibinfo{author}{\bibfnamefont{S.}~\bibnamefont{Kahkonen}},
  \bibinfo{journal}{Neuroscience and Biobehavioral Reviews}
  \textbf{\bibinfo{volume}{28}}, \bibinfo{pages}{827} (\bibinfo{year}{2005}).

\bibitem[{\citenamefont{Friston et~al.}(1993)\citenamefont{Friston, Frith,
  Liddle, and Frackowiak}}]{friston1993}
\bibinfo{author}{\bibfnamefont{K.~J.} \bibnamefont{Friston}},
  \bibinfo{author}{\bibfnamefont{C.~D.} \bibnamefont{Frith}},
  \bibinfo{author}{\bibfnamefont{P.~F.} \bibnamefont{Liddle}},
  \bibnamefont{and} \bibinfo{author}{\bibfnamefont{R.~S.~J.}
  \bibnamefont{Frackowiak}}, \bibinfo{journal}{Journal of Cerebral Blood Flow
  and Metabolism} \textbf{\bibinfo{volume}{13}}, \bibinfo{pages}{5}
  (\bibinfo{year}{1993}).

\bibitem[{\citenamefont{Hebb}(1949)}]{hebb1949}
\bibinfo{author}{\bibfnamefont{D.}~\bibnamefont{Hebb}},
  \emph{\bibinfo{title}{The Organization of Behavior}}
  (\bibinfo{publisher}{Wiley}, \bibinfo{address}{New York},
  \bibinfo{year}{1949}).

\bibitem[{\citenamefont{Gerstein et~al.}(1978)\citenamefont{Gerstein, Perkel,
  and Subramanian}}]{gerstein1978}
\bibinfo{author}{\bibfnamefont{G.~L.} \bibnamefont{Gerstein}},
  \bibinfo{author}{\bibfnamefont{D.~H.} \bibnamefont{Perkel}},
  \bibnamefont{and} \bibinfo{author}{\bibfnamefont{K.~N.}
  \bibnamefont{Subramanian}}, \bibinfo{journal}{Brain Res}
  \textbf{\bibinfo{volume}{140}}, \bibinfo{pages}{43} (\bibinfo{year}{1978}).

\bibitem[{\citenamefont{Singer}(1999)}]{singer1999}
\bibinfo{author}{\bibfnamefont{W.}~\bibnamefont{Singer}},
  \bibinfo{journal}{Neuron} \textbf{\bibinfo{volume}{24}}, \bibinfo{pages}{49}
  (\bibinfo{year}{1999}).

\bibitem[{\citenamefont{Zhou et~al.}(2007)\citenamefont{Zhou, Zemanova,
  Zamora-Lopez, Hilgetag, and Kurths}}]{zhou2007}
\bibinfo{author}{\bibfnamefont{C.~S.} \bibnamefont{Zhou}},
  \bibinfo{author}{\bibfnamefont{L.}~\bibnamefont{Zemanova}},
  \bibinfo{author}{\bibfnamefont{G.}~\bibnamefont{Zamora-Lopez}},
  \bibinfo{author}{\bibfnamefont{C.~C.} \bibnamefont{Hilgetag}},
  \bibnamefont{and} \bibinfo{author}{\bibfnamefont{J.}~\bibnamefont{Kurths}},
  \bibinfo{journal}{New Journal of Physics} \textbf{\bibinfo{volume}{9}},
  (\bibinfo{year}{2007}).

\bibitem[{\citenamefont{Milner}(1974)}]{milner1974}
\bibinfo{author}{\bibfnamefont{P.~M.} \bibnamefont{Milner}},
  \bibinfo{journal}{Psychol Rev} \textbf{\bibinfo{volume}{81}},
  \bibinfo{pages}{521} (\bibinfo{year}{1974}).

\bibitem[{\citenamefont{Engel and Singer}(2001)}]{engel2001}
\bibinfo{author}{\bibfnamefont{A.~K.} \bibnamefont{Engel}} \bibnamefont{and}
  \bibinfo{author}{\bibfnamefont{W.}~\bibnamefont{Singer}},
  \bibinfo{journal}{Trends Cogn Sci} \textbf{\bibinfo{volume}{5}},
  \bibinfo{pages}{16} (\bibinfo{year}{2001}).

\bibitem[{\citenamefont{von~der Malsburg}(1995)}]{vondermalsburg1995}
\bibinfo{author}{\bibfnamefont{C.}~\bibnamefont{von~der Malsburg}},
  \bibinfo{journal}{Curr Opin Neurobiol} \textbf{\bibinfo{volume}{5}},
  \bibinfo{pages}{520} (\bibinfo{year}{1995}).

\bibitem[{\citenamefont{Singer}(2001)}]{singer2001}
\bibinfo{author}{\bibfnamefont{W.}~\bibnamefont{Singer}}, \bibinfo{journal}{Ann
  N Y Acad Sci} \textbf{\bibinfo{volume}{929}}, \bibinfo{pages}{123}
  (\bibinfo{year}{2001}).

\bibitem[{\citenamefont{von~der Malsburg}(1981)}]{vondermalsburg1981}
\bibinfo{author}{\bibfnamefont{C.}~\bibnamefont{von~der Malsburg}},
  \bibinfo{type}{Internal Report}, \bibinfo{institution}{MPI Biophysical
  Chemistry} (\bibinfo{year}{1981}).

\bibitem[{\citenamefont{Gray}(1999)}]{gray1999}
\bibinfo{author}{\bibfnamefont{C.~M.} \bibnamefont{Gray}},
  \bibinfo{journal}{Neuron} \textbf{\bibinfo{volume}{24}}, \bibinfo{pages}{31}
  (\bibinfo{year}{1999}).

\bibitem[{\citenamefont{Singer}(1993)}]{singer1993}
\bibinfo{author}{\bibfnamefont{W.}~\bibnamefont{Singer}},
  \bibinfo{journal}{Annu Rev Physiol} \textbf{\bibinfo{volume}{55}},
  \bibinfo{pages}{349} (\bibinfo{year}{1993}).

\bibitem[{\citenamefont{Engel et~al.}(1991)\citenamefont{Engel, Konig, and
  Singer}}]{engel1991}
\bibinfo{author}{\bibfnamefont{A.~K.} \bibnamefont{Engel}},
  \bibinfo{author}{\bibfnamefont{P.}~\bibnamefont{Konig}}, \bibnamefont{and}
  \bibinfo{author}{\bibfnamefont{W.}~\bibnamefont{Singer}},
  \bibinfo{journal}{Proc Natl Acad Sci U S A} \textbf{\bibinfo{volume}{88}},
  \bibinfo{pages}{9136} (\bibinfo{year}{1991}).

\bibitem[{\citenamefont{Gerstein et~al.}(1985)\citenamefont{Gerstein, Perkel,
  and Dayhoff}}]{gerstein1985}
\bibinfo{author}{\bibfnamefont{G.~L.} \bibnamefont{Gerstein}},
  \bibinfo{author}{\bibfnamefont{D.~H.} \bibnamefont{Perkel}},
  \bibnamefont{and} \bibinfo{author}{\bibfnamefont{J.~E.}
  \bibnamefont{Dayhoff}}, \bibinfo{journal}{J Neurosci}
  \textbf{\bibinfo{volume}{5}}, \bibinfo{pages}{881} (\bibinfo{year}{1985}).

\bibitem[{\citenamefont{Baker and Gerstein}(2000)}]{baker2000}
\bibinfo{author}{\bibfnamefont{S.~N.} \bibnamefont{Baker}} \bibnamefont{and}
  \bibinfo{author}{\bibfnamefont{G.~L.} \bibnamefont{Gerstein}},
  \bibinfo{journal}{Neural Comput} \textbf{\bibinfo{volume}{12}},
  \bibinfo{pages}{2597} (\bibinfo{year}{2000}).

\bibitem[{\citenamefont{Dayhoff}(1994)}]{dayhoff1994}
\bibinfo{author}{\bibfnamefont{J.~E.} \bibnamefont{Dayhoff}},
  \bibinfo{journal}{Biol Cybern} \textbf{\bibinfo{volume}{71}},
  \bibinfo{pages}{263} (\bibinfo{year}{1994}).

\bibitem[{\citenamefont{Lindsey and Gerstein}(2006)}]{lindsey2006}
\bibinfo{author}{\bibfnamefont{B.~G.} \bibnamefont{Lindsey}} \bibnamefont{and}
  \bibinfo{author}{\bibfnamefont{G.~L.} \bibnamefont{Gerstein}},
  \bibinfo{journal}{J Neurosci Methods} \textbf{\bibinfo{volume}{150}},
  \bibinfo{pages}{116} (\bibinfo{year}{2006}).

\bibitem[{\citenamefont{Borgatti}(1994)}]{borgatti1994}
\bibinfo{author}{\bibfnamefont{S.}~\bibnamefont{Borgatti}},
  \bibinfo{journal}{Connections} \textbf{\bibinfo{volume}{17}},
  \bibinfo{pages}{78} (\bibinfo{year}{1994}).

\bibitem[{\citenamefont{Boccaletti et~al.}(2006)\citenamefont{Boccaletti,
  Latora, Moreno, Chavez, and Hwang}}]{boccaletti2006}
\bibinfo{author}{\bibfnamefont{S.}~\bibnamefont{Boccaletti}},
  \bibinfo{author}{\bibfnamefont{V.}~\bibnamefont{Latora}},
  \bibinfo{author}{\bibfnamefont{Y.}~\bibnamefont{Moreno}},
  \bibinfo{author}{\bibfnamefont{M.}~\bibnamefont{Chavez}}, \bibnamefont{and}
  \bibinfo{author}{\bibfnamefont{D.~U.} \bibnamefont{Hwang}},
  \bibinfo{journal}{Physics Reports} \textbf{\bibinfo{volume}{424}},
  \bibinfo{pages}{175} (\bibinfo{year}{2006}).

\bibitem[{\citenamefont{Berger et~al.}(2007)\citenamefont{Berger, Warren,
  Normann, Arieli, and Grun}}]{berger2007}
\bibinfo{author}{\bibfnamefont{D.}~\bibnamefont{Berger}},
  \bibinfo{author}{\bibfnamefont{D.}~\bibnamefont{Warren}},
  \bibinfo{author}{\bibfnamefont{R.}~\bibnamefont{Normann}},
  \bibinfo{author}{\bibfnamefont{A.}~\bibnamefont{Arieli}}, \bibnamefont{and}
  \bibinfo{author}{\bibfnamefont{S.}~\bibnamefont{Grun}},
  \bibinfo{journal}{Neurocomputing} \textbf{\bibinfo{volume}{70}},
  \bibinfo{pages}{2112} (\bibinfo{year}{2007}).

\bibitem[{\citenamefont{Kreuz et~al.}(2007)\citenamefont{Kreuz, Haas, Morelli,
  Abarbanel, and Politi}}]{kreuz2007}
\bibinfo{author}{\bibfnamefont{T.}~\bibnamefont{Kreuz}},
  \bibinfo{author}{\bibfnamefont{J.~S.} \bibnamefont{Haas}},
  \bibinfo{author}{\bibfnamefont{A.}~\bibnamefont{Morelli}},
  \bibinfo{author}{\bibfnamefont{H.~D.~I.} \bibnamefont{Abarbanel}},
  \bibnamefont{and} \bibinfo{author}{\bibfnamefont{A.}~\bibnamefont{Politi}},
  \bibinfo{journal}{Journal of Neuroscience Methods}
  \textbf{\bibinfo{volume}{165}}, \bibinfo{pages}{151} (\bibinfo{year}{2007}).

\bibitem[{\citenamefont{Ozden et~al.}(2008)\citenamefont{Ozden, Lee, Sullivan,
  and Wang}}]{ozden2008}
\bibinfo{author}{\bibfnamefont{I.}~\bibnamefont{Ozden}},
  \bibinfo{author}{\bibfnamefont{H.~M.} \bibnamefont{Lee}},
  \bibinfo{author}{\bibfnamefont{M.~R.} \bibnamefont{Sullivan}},
  \bibnamefont{and} \bibinfo{author}{\bibfnamefont{S.~S.~H.}
  \bibnamefont{Wang}}, \bibinfo{journal}{Journal of Neurophysiology}
  \textbf{\bibinfo{volume}{100}}, \bibinfo{pages}{495} (\bibinfo{year}{2008}).

\bibitem[{\citenamefont{Newman and Girvan}(2004)}]{newman2004a}
\bibinfo{author}{\bibfnamefont{M.~E.} \bibnamefont{Newman}} \bibnamefont{and}
  \bibinfo{author}{\bibfnamefont{M.}~\bibnamefont{Girvan}},
  \bibinfo{journal}{Phys. Rev. E} \textbf{\bibinfo{volume}{69}},
  \bibinfo{pages}{026113} (\bibinfo{year}{2004}).

\bibitem[{\citenamefont{Newman}(2004{\natexlab{a}})}]{newman2004b}
\bibinfo{author}{\bibfnamefont{M.}~\bibnamefont{Newman}},
  \bibinfo{journal}{Phys. Rev. E} \textbf{\bibinfo{volume}{70}},
  \bibinfo{pages}{056131} (\bibinfo{year}{2004}{\natexlab{a}}).

\bibitem[{\citenamefont{Fortunato et~al.}(2004)\citenamefont{Fortunato, Latora,
  and Marchiori}}]{fortunato2004}
\bibinfo{author}{\bibfnamefont{S.}~\bibnamefont{Fortunato}},
  \bibinfo{author}{\bibfnamefont{V.}~\bibnamefont{Latora}}, \bibnamefont{and}
  \bibinfo{author}{\bibfnamefont{M.}~\bibnamefont{Marchiori}},
  \bibinfo{journal}{Phys. Rev. E} \textbf{\bibinfo{volume}{70}},
  \bibinfo{pages}{056104} (\bibinfo{year}{2004}).

\bibitem[{\citenamefont{Zhou}(2003)}]{zhou2003}
\bibinfo{author}{\bibfnamefont{H.}~\bibnamefont{Zhou}}, \bibinfo{journal}{Phys
  Rev E} \textbf{\bibinfo{volume}{67}}, \bibinfo{pages}{041908}
  (\bibinfo{year}{2003}).

\bibitem[{\citenamefont{Newman}(2004{\natexlab{b}})}]{newman2004c}
\bibinfo{author}{\bibfnamefont{M.~E.} \bibnamefont{Newman}},
  \bibinfo{journal}{Phys Rev E} \textbf{\bibinfo{volume}{69}},
  \bibinfo{pages}{066133} (\bibinfo{year}{2004}{\natexlab{b}}).

\bibitem[{\citenamefont{Ball and Hall}(1967)}]{ball1967}
\bibinfo{author}{\bibfnamefont{G.~H.} \bibnamefont{Ball}} \bibnamefont{and}
  \bibinfo{author}{\bibfnamefont{D.~J.} \bibnamefont{Hall}},
  \bibinfo{journal}{Behav Sci} \textbf{\bibinfo{volume}{12}},
  \bibinfo{pages}{153} (\bibinfo{year}{1967}).

\bibitem[{\citenamefont{Rolston et~al.}(2007)\citenamefont{Rolston, Wagenaar,
  and Potter}}]{rolston2007}
\bibinfo{author}{\bibfnamefont{J.~D.} \bibnamefont{Rolston}},
  \bibinfo{author}{\bibfnamefont{D.~A.} \bibnamefont{Wagenaar}},
  \bibnamefont{and} \bibinfo{author}{\bibfnamefont{S.~M.}
  \bibnamefont{Potter}}, \bibinfo{journal}{Neuroscience}
  \textbf{\bibinfo{volume}{148}}, \bibinfo{pages}{294} (\bibinfo{year}{2007}).

\bibitem[{\citenamefont{Date et~al.}(1998)\citenamefont{Date, Bienenstock, and
  Geman}}]{date1998}
\bibinfo{author}{\bibfnamefont{A.}~\bibnamefont{Date}},
  \bibinfo{author}{\bibfnamefont{E.}~\bibnamefont{Bienenstock}},
  \bibnamefont{and} \bibinfo{author}{\bibfnamefont{S.}~\bibnamefont{Geman}},
  \bibinfo{type}{Technical Report}, \bibinfo{institution}{Division of Applied
  Mathematics, Brown University} (\bibinfo{year}{1998}).

\bibitem[{\citenamefont{Shmiel et~al.}(2006)\citenamefont{Shmiel, Drori,
  Shmiel, Ben-Shaul, Nadasdy, Shemesh, Teicher, and Abeles}}]{shmiel2006}
\bibinfo{author}{\bibfnamefont{T.}~\bibnamefont{Shmiel}},
  \bibinfo{author}{\bibfnamefont{R.}~\bibnamefont{Drori}},
  \bibinfo{author}{\bibfnamefont{O.}~\bibnamefont{Shmiel}},
  \bibinfo{author}{\bibfnamefont{Y.}~\bibnamefont{Ben-Shaul}},
  \bibinfo{author}{\bibfnamefont{Z.}~\bibnamefont{Nadasdy}},
  \bibinfo{author}{\bibfnamefont{M.}~\bibnamefont{Shemesh}},
  \bibinfo{author}{\bibfnamefont{M.}~\bibnamefont{Teicher}}, \bibnamefont{and}
  \bibinfo{author}{\bibfnamefont{M.}~\bibnamefont{Abeles}},
  \bibinfo{journal}{Journal of Neurophysiology} \textbf{\bibinfo{volume}{96}},
  \bibinfo{pages}{2645} (\bibinfo{year}{2006}).

\bibitem[{\citenamefont{Pazienti et~al.}(2007)\citenamefont{Pazienti, Diesmann,
  and Grun}}]{pazienti2007}
\bibinfo{author}{\bibfnamefont{A.}~\bibnamefont{Pazienti}},
  \bibinfo{author}{\bibfnamefont{M.}~\bibnamefont{Diesmann}}, \bibnamefont{and}
  \bibinfo{author}{\bibfnamefont{S.}~\bibnamefont{Grun}},
  \bibinfo{journal}{Advances in Brain, Vision, and Artificial Intelligence,
  Proceedings} \textbf{\bibinfo{volume}{4729}}, \bibinfo{pages}{428}
  (\bibinfo{year}{2007}).

\bibitem[{\citenamefont{Fred and Jain}(2003)}]{fred2003}
\bibinfo{author}{\bibfnamefont{A.~L.~N.} \bibnamefont{Fred}} \bibnamefont{and}
  \bibinfo{author}{\bibfnamefont{A.~K.} \bibnamefont{Jain}},
  \bibinfo{journal}{2003 IEEE Computer Society Conference on Computer Vision
  and Pattern Recognition, Vol II, Proceedings} pp. \bibinfo{pages}{128--133}
  (\bibinfo{year}{2003}).

\bibitem[{\citenamefont{Berke et~al.}(2008)\citenamefont{Berke, Hetrick, Breck,
  and Greene}}]{berke2008}
\bibinfo{author}{\bibfnamefont{J.~D.} \bibnamefont{Berke}},
  \bibinfo{author}{\bibfnamefont{V.}~\bibnamefont{Hetrick}},
  \bibinfo{author}{\bibfnamefont{J.}~\bibnamefont{Breck}}, \bibnamefont{and}
  \bibinfo{author}{\bibfnamefont{R.~W.} \bibnamefont{Greene}},
  \bibinfo{journal}{Hippocampus} \textbf{\bibinfo{volume}{18}},
  \bibinfo{pages}{519} (\bibinfo{year}{2008}).

\bibitem[{\citenamefont{Buzsaki et~al.}(2003)\citenamefont{Buzsaki, Buhl,
  Harris, Csicsvari, Czeh, and Morozov}}]{buzsaki2003}
\bibinfo{author}{\bibfnamefont{G.}~\bibnamefont{Buzsaki}},
  \bibinfo{author}{\bibfnamefont{D.~L.} \bibnamefont{Buhl}},
  \bibinfo{author}{\bibfnamefont{K.~D.} \bibnamefont{Harris}},
  \bibinfo{author}{\bibfnamefont{J.}~\bibnamefont{Csicsvari}},
  \bibinfo{author}{\bibfnamefont{B.}~\bibnamefont{Czeh}}, \bibnamefont{and}
  \bibinfo{author}{\bibfnamefont{A.}~\bibnamefont{Morozov}},
  \bibinfo{journal}{Neuroscience} \textbf{\bibinfo{volume}{116}},
  \bibinfo{pages}{201} (\bibinfo{year}{2003}).

\bibitem[{\citenamefont{Harris et~al.}(2003)\citenamefont{Harris, Csicsvari,
  Hirase, Dragoi, and Buzsaki}}]{harris2003}
\bibinfo{author}{\bibfnamefont{K.~D.} \bibnamefont{Harris}},
  \bibinfo{author}{\bibfnamefont{J.}~\bibnamefont{Csicsvari}},
  \bibinfo{author}{\bibfnamefont{H.}~\bibnamefont{Hirase}},
  \bibinfo{author}{\bibfnamefont{G.}~\bibnamefont{Dragoi}}, \bibnamefont{and}
  \bibinfo{author}{\bibfnamefont{G.}~\bibnamefont{Buzsaki}},
  \bibinfo{journal}{Nature} \textbf{\bibinfo{volume}{424}},
  \bibinfo{pages}{552} (\bibinfo{year}{2003}).

\bibitem[{\citenamefont{Foster and Wilson}(2006)}]{foster2006}
\bibinfo{author}{\bibfnamefont{D.~J.} \bibnamefont{Foster}} \bibnamefont{and}
  \bibinfo{author}{\bibfnamefont{M.~A.} \bibnamefont{Wilson}},
  \bibinfo{journal}{Nature} \textbf{\bibinfo{volume}{440}},
  \bibinfo{pages}{680} (\bibinfo{year}{2006}).

\bibitem[{\citenamefont{Buzsaki}(1998)}]{buzsaki1998}
\bibinfo{author}{\bibfnamefont{G.}~\bibnamefont{Buzsaki}}, \bibinfo{journal}{J
  Sleep Res} \textbf{\bibinfo{volume}{7 Suppl 1}}, \bibinfo{pages}{17}
  (\bibinfo{year}{1998}).

\bibitem[{\citenamefont{Kudrimoti et~al.}(1999)\citenamefont{Kudrimoti, Barnes,
  and McNaughton}}]{kudrimoti1999}
\bibinfo{author}{\bibfnamefont{H.~S.} \bibnamefont{Kudrimoti}},
  \bibinfo{author}{\bibfnamefont{C.~A.} \bibnamefont{Barnes}},
  \bibnamefont{and} \bibinfo{author}{\bibfnamefont{B.~L.}
  \bibnamefont{McNaughton}}, \bibinfo{journal}{J Neurosci}
  \textbf{\bibinfo{volume}{19}}, \bibinfo{pages}{4090} (\bibinfo{year}{1999}).

\bibitem[{\citenamefont{O'Neill et~al.}(2008)\citenamefont{O'Neill, Senior,
  Allen, Huxter, and Csicsvari}}]{oneill2008}
\bibinfo{author}{\bibfnamefont{J.}~\bibnamefont{O'Neill}},
  \bibinfo{author}{\bibfnamefont{T.~J.} \bibnamefont{Senior}},
  \bibinfo{author}{\bibfnamefont{K.}~\bibnamefont{Allen}},
  \bibinfo{author}{\bibfnamefont{J.~R.} \bibnamefont{Huxter}},
  \bibnamefont{and}
  \bibinfo{author}{\bibfnamefont{J.}~\bibnamefont{Csicsvari}},
  \bibinfo{journal}{Nat Neurosci} \textbf{\bibinfo{volume}{11}},
  \bibinfo{pages}{209} (\bibinfo{year}{2008}).

\end{thebibliography}

\end{document}